\documentclass{Interspeech}

\interspeechcameraready

\title{Explainable Speech Emotion Recognition Through Attentive Pooling: Insights~from Attention-Based Temporal Localization}

\author{Tahitoa}{Leygue}
\author{Astrid}{Sabourin}
\author{Christian}{Bolzmacher}
\author{Sylvain}{Bouchigny}
\author{Margarita}{Anastassova}
\author{Quoc-Cuong}{Pham}

\affiliation[nocounter]{Université Paris-Saclay}{CEA, List, F-91120, Palaiseau}{France}

\email{tahitoa.leygue@cea.fr, quoc-cuong.pham@cea.fr}

\usepackage{comment}

\begin{document}

\maketitle

\begin{abstract}
State-of-the-art transformer models for Speech Emotion Recognition (SER) rely on temporal feature aggregation, yet advanced pooling methods remain underexplored. We systematically benchmark pooling strategies, including Multi-Query Multi-Head Attentive Statistics Pooling, which achieves a 3.5 percentage point macro F1 gain over average pooling. Attention analysis shows 15 percent of frames capture 80 percent of emotion cues, revealing a localized pattern of emotional information. Analysis of high-attention frames reveals that non-linguistic vocalizations and hyperarticulated phonemes are disproportionately prioritized during pooling, mirroring human perceptual strategies. Our findings position attentive pooling as both a performant SER mechanism and a biologically plausible tool for explainable emotion localization. On Interspeech 2025 Speech Emotion Recognition in Naturalistic Conditions Challenge, our approach obtained a macro F1 score of 0.3649.
\end{abstract}

\section{Introduction}

Speech serves as humanity's primary communication medium, conveying both linguistic content and paralinguistic cues about emotions, intentions, and context.
The complexities of vocal modulations (pitch, speech rate, variance) provide crucial insights into speakers' emotional states~\cite{KhalilSpeechEmotion19}.
Emotions---complex psychological and physiological responses to stimuli---are fundamental to human communication and social interaction~\cite{AkcaySpeechemotion2001}.

Traditional SER systems relied on handcrafted acoustic features---combining prosodic elements (pitch, energy, duration) and spectral descriptors (MFCCs, formants)~\cite{SchullerSpeechemotion04a}---and used conventional classifiers such as SVMs~\cite{ChavhanSpeech10}, HMMs~\cite{NweSpeechemotion0311a}, and GMMs~\cite{ReynoldsRobusttextindependent9501}.
While effective for acted datasets, these approaches underperformed with spontaneous emotions in real-world scenarios~\cite{SchullerINTERSPEECH20090909}.
Early integration of basic textual features~\cite{ChulMinLeedetectingemotions0503} was also limited in capturing complex semantic relationships.

Deep learning has revolutionized SER with end-to-end architectures that automatically learn features.
Early innovations used CNNs for spectrogram analysis \cite{TrigeorgisAdieufeatures1603} and attention-enhanced LSTMs \cite{MirsamadiAutomaticspeech1703}, leading to advanced models like Wav2Vec2.0 \cite{Baevskiwa2vec220} and HuBERT \cite{HsuHuBERTSelfSupervised21}.
Originally for speech recognition, these models now excel in various audio tasks including SER by learning powerful representations from unlabeled data.
Recent developments like W2V-BERT 2.0 (600M parameters, trained on 4.5M hours of audio) \cite{barrault2023seamlessm4t} further underscore their potential for SER tasks.

For textual analysis, BERT-based architectures~\cite{kenton2019bert} excel at capturing contextual information, making them a popular choice for tasks that involve processing text derived from audio transcripts.
Compared to large language models (LLMs) such as LLaMA~\cite{GrattafioriLlama32411}, BERT offers a compelling trade-off between performance and computational efficiency.

Multimodal approaches which combine audio representations with their corresponding textual transcripts have shown improved robustness~\cite{XuLearningAlignment19}, though determining optimal strategies for aggregating these features remains an open area of research.
The integration of lexical information alongside acoustic features has consistently demonstrated superior performance in emotion recognition tasks compared to unimodal approaches.
Recent architectures leverage pre-trained models for both modalities, such as Wav2Vec 2.0 for speech and BERT variants or LLMs for text, thereby benefiting from transfer learning.
These approaches have proven particularly effective in the Odyssey 2024 Challenge~\cite{GoncalvesOdyssey20242406}, where multimodal systems consistently outperformed their unimodal counterparts~\cite{Chen1stPlace2405b, HarmTalTechSystems2406}.

The transition from frame-level to utterance-level representations is a pivotal step in SER architectures, particularly in the era of self-supervised speech models.
Models like Wav2Vec 2.0 have proven adept at extracting rich frame-wise features, but the challenge lies in aggregating these fine-grained temporal representations into effective utterance-level embeddings.
While the temporal dynamics of speech are inherently complex, the choice of aggregation mechanisms significantly impacts the performance of downstream emotion recognition tasks.

Historically, simple pooling methods like mean, max, and downsampling gained popularity from text embeddings~\cite{MohamedSelfSupervisedSpeech2210}, but struggled with speech's complex temporal dependencies.
Statistics Pooling~\cite{SnyderDeepNeural1708} advanced this by incorporating higher-order moments, capturing richer patterns crucial for self-supervised SER.

A significant advancement came with Attentive Statistics~(AS) Pooling~\cite{Okabe_2018}, which introduced learnable parameters to dynamically weight frames, enabling better capture of long-range dependencies.
This innovation led to Multi-Head variants including Self-Attentive~(SA) pooling~\cite{zhu2018self} and Multi-Head Attention~(MHA) Pooling~\cite{IndiaSelfMultiHead1907}.
The most recent advancement, Multi-Query Multi-Head Attentive~(MQMHA) Pooling~\cite{ZhaoMultiQueryMultiHead2205}, can be viewed as a generalization of these approaches, combining multiple attention mechanisms with statistical aggregation.

Speaker identification determines who is speaking, while SER analyzes emotional content in speech.
Both examine vocal characteristics, with MQMHA effectively capturing subtle acoustic features for each.
For identification, attention heads focus on consistent traits like timbre; for SER, they track dynamic features such as prosody and speech rate indicating emotions.
Both tasks benefit from MQMHA's detailed analysis while targeting different speech characteristics.

We present a multimodal framework for SER that leverages the MSP-Podcast corpus dataset~\cite{LotfianBuildingNaturalistic1910b}.
Developed for the Task~1 on the Interspeech~2025 Speech Emotion Recognition in Naturalistic Conditions Challenge, our framework introduces three key innovations.
We first introduce a flexible architecture paired with a training procedure that gradually unfreezes pretrained audio and text encoders.
This approach enables the integration of diverse pooling strategies, mitigates overfitting, and ensures computational efficiency.
Secondly, we systematically evaluate various pooling mechanisms and adopt MQMHA Pooling, which consistently improves performance on the speech emotion recognition task.
Finally, through attention weight analysis, we provide empirical evidence that emotional markers are temporally localized in speech, with attention patterns aligning with specific speech characteristics such as emphasized syllables.
These findings advance our understanding of how temporal dynamics contribute to emotion recognition in naturalistic conditions.

\section{Methodology}

\subsection{Model architecture}

\subsubsection{Audio and Text Encoders}

\begin{figure}[b]
  \centering
  \includegraphics[width=0.7\linewidth]{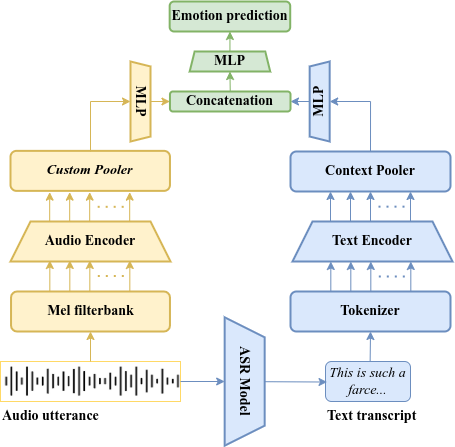}
  \caption{Multimodal model architecture. ``Custom Pooler" denotes one of the pooling strategy described in Section~\ref{sec:pooling_strategies}}
  \label{fig:model_architecture}
\end{figure} 

Our architecture leverages state-of-the-art pretrained models for both audio and text processing, is represented in Figure~\ref{fig:model_architecture}.
For audio encoding, we employ W2V-BERT 2.0~\cite{barrault2023seamlessm4t}, which has demonstrated superior performance in paralinguistic tasks through its self-supervised pretraining on large-scale speech data.
We use the pretrained checkpoint available at \url{https://huggingface.co/facebook/w2v-bert-2.0}.

For text encoding, we utilize DeBERTa v3~\cite{HeDeBERTaV3Improving2303a}, which offers several advantages over traditional BERT-based models for our emotion recognition task.
The model's disentangled attention mechanism and enhanced positional encoding improve the capture of nuanced emotional content in transcribed speech by representing each word with two distinct vectors: one for position and one for content.
While large language models like LLaMA 3~\cite{GrattafioriLlama32411} have shown impressive results in various NLP tasks, we opted for DeBERTa v3 due to its computational efficiency and proven effectiveness in emotion-specific tasks~\cite{ImranEmotionClassification2404}.
The pretrained model is available at \url{https://huggingface.co/microsoft/deberta-v3-base}.

\subsubsection{Multimodal Fusion}

The fusion module concatenates the pooled embeddings from both modalities, followed by a Multi-Layer Perceptron (MLP) classifier.
This approach, while simple, has shown robust performance in multimodal emotion recognition tasks and during Odyssey 2024 challenge~\cite{GoncalvesOdyssey20242406, Chen1stPlace2405b, HarmTalTechSystems2406}.

\subsection{Pooling Strategies}
\label{sec:pooling_strategies}

\begin{figure}[t]
  \centering
  \includegraphics[width=\linewidth]{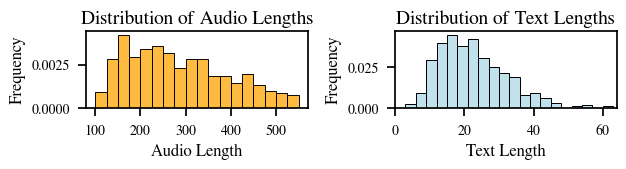}
  \caption{Distribution of the number of frames per dataset item for audio and text modalities.}
  \label{fig:features_lenghts}
\end{figure}

A key insight from our analysis of the dataset highlights the significant disparity in length between audio data and its text transcriptions. Specifically, audio sequences contain a substantially higher number of frames compared to the token count in their corresponding textual representations, as illustrated by the histograms of data lengths in Figure~\ref{fig:features_lenghts}. This stark difference underscores the unique challenges posed by the high temporal resolution of audio data, motivating our investigation into advanced pooling strategies for effectively aggregating these fine-grained frame-level features into robust utterance-level representations.

\subsubsection{Basic Pooling Methods}

To ensure a transparent basis for evaluating the performance of our custom pooling layers, we implement standard aggregation methods that account for masking to handle variable-length sequences.
This masking mechanism ensures robustness by avoiding any influence from padding tokens, thereby eliminating sources of error.

Each method operates on a batch of input feature matrices $\mathbf{X} \in \mathbb{R}^{B \times T \times K}$, where $B$ is the batch size, $T$ the maximum sequence length, and $K$ the feature size, with corresponding masking matrices $\mathbf{M} \in \{0,1\}^{B \times T}$ to denote valid frames.

\begin{enumerate}
    \item Max Pooling can be expressed as $\mathbf{Y} \in \mathbb{R}^{B \times K}$ where:
    \begin{equation}
        \mathbf{Y}_{b} = \max_{t \in \{1, \dots, T\}} \big(\mathbf{X}_{b,t} \cdot {M}_{b,t}\big)
    \end{equation}
    \item Average Pooling can be expressed as $\mathbf{Y} \in \mathbb{R}^{B \times K}$ where:
    \begin{equation}
        \mathbf{Y}_{b} = \boldsymbol{\mu}_{b} = \frac{\sum_{t=1}^{T} \mathbf{X}_{b,t} \cdot {M}_{b,t}}{\sum_{t=1}^{T} {M}_{b,t}}
    \end{equation}
    \item Statistics Pooling is $\mathbf{Y} \in \mathbb{R}^{B \times 2K}$ where $\mathbf{Y}_{b} = \left[\boldsymbol{\mu}_{b}, \boldsymbol{\sigma}_{b}\right]$ and $(\cdot)^{\odot 2}$ denotes the element-wise squaring:
\end{enumerate}

\begin{equation}
\boldsymbol{\sigma}_{b} = \sqrt{\frac{\sum_{t=1}^{T} {M}_{b,t} \cdot \left(\mathbf{X}_{b,t} - \boldsymbol{\mu}_{b}\right)^{\odot 2}}{\sum_{t=1}^{T} {M}_{b,t}}}
\end{equation}

\subsubsection{Advanced Pooling strategies}

We implement MQMHA Pooling~\cite{ZhaoMultiQueryMultiHead2205}, which generalizes existing attention-based pooling methods.
Using the same notations as above, we also introduce $Q \in \mathbb{N}, H \in \mathbb{N}$ respectively the number of queries and heads.

Let $K'=K/H \in \mathbb{N}$, we partition $\mathbf{X}$ into $H$ equal parts such that $\mathbf{X} = [\mathbf{X}^{(1)}, \mathbf{X}^{(2)}, \dots, \mathbf{X}^{(H)}], \quad \mathbf{X}^{(h)} \in \mathbb{R}^{B\times T\times K'}$.
Let $F_{n,p}^{(q,h)}$ denote either a linear mapping (for $n=1$) or a two-layer MLP with a hidden layer of dimension $p$ and a ReLU activation (for $n=2$); $q$ and $h$ are the query and head indices, respectively.
We sequentially apply the $F_{n,p}^{(q,h)}(\cdot)$ and discard masked layers:
\begin{equation} 
    {e}_{b,t}^{(q,h)} = F_{n,p}^{(q)}\left(\mathbf{X}_{b,t}^{(h)}\right) \label{eq:1}
\end{equation}

\begin{equation} 
    {\tilde{e}}_{b,t}^{(q,h)} = \begin{cases}
            e_{b,t}^{(q,h)} & \text{if } M_{b,t} = 1 \\
            -\infty & \text{otherwise}
        \end{cases} \label{eq:2}
\end{equation}

We finally apply softmax to get the attention weights, with ${\omega}_{b,t}^{(q,h)}$ being the weight associated to the score ${\tilde{e}}_{b,t}^{(q,h)}$.

Representation of weighted mean and standard deviation can be computed from the following equations.

\begin{equation}
    \boldsymbol{\mu}_b^{(q,h)} = \sum_{t=1}^T \omega_{b,t}^{(q,h)} \mathbf{X}_{b,t}^{(h)}
\end{equation}
\begin{equation}
    \boldsymbol{\sigma}_b^{(q,h)} = \sqrt{\sum_{t=1}^T \omega_{b,t}^{(q,h)} \left(\mathbf{X}_{b,t}^{(h)}\right)^{\odot 2} - \left(\boldsymbol{\mu}_b^{(q,h)}\right)^{\odot 2}}
\end{equation}

The output $\mathbf{Y} \in \mathbb{R}^{B\times 2QK}$ is obtained by concatenating the outputs from all queries and heads. This formulation encompasses several existing attention mechanisms as special cases.

\subsection{Training protocol}

Our training strategy employs gradual unfreezing, a technique that balances the stability-flexibility trade-off in transfer learning~\cite{HowardUniversalLanguage1805a}.
Fully unfreezing all layers initially risks catastrophic forgetting (where over-parameterized models overwrite pretrained knowledge and overfit to sparse emotion labels), while freezing all but the last layers limits adaptation capacity.
Gradual unfreezing navigates this compromise by incrementally exposing pretrained parameters to task-specific features, enabling controlled evolution of multimodal representations.
This preserves generalizable acoustic-linguistic patterns while adapting to emotional semantics, facilitating robust convergence through curriculum-inspired parameter updates.

Our protocol consists of three phases.
First, we freeze both encoders and train only the classification head. This initial phase allows the fusion layer to learn cross-modal relationships while preserving the pretrained representations. Subsequently, we selectively unfreeze the upper transformer layers of both encoders, following the intuition that higher layers capture more task-specific features. Finally, we unfreeze all parameters, including the embedding layers and feature extractors. Transition between steps are triggered by an early stopping module.

To address class imbalance we employ a class-weighted focal loss:

\begin{equation}
    \mathcal{L} = -\alpha_c(1-p_c)^\gamma \log(p_c)
\end{equation}

where $p_c$ is the model's estimated probability for the target class, $\alpha_c$ is the class weight computed as the inverse of class frequency, and $\gamma=2$.

For optimization, we use AdamW with weight decay $\lambda=\nolinebreak0.01$ and implement a linear learning rate warm-up followed by cosine decay.
The learning rates are scaled progressively across the phases:  $\eta_1 = 1 \times 10^{-5}$  in phase 1,  $\eta_2 = 3 \times 10^{-6}$  in phase 2, and  $\eta_3 = 1 \times 10^{-6}$  in phase 3.

\section{Experimental setup}

In this section, we outline our experimental design, detailing the dataset and evaluation protocol.
We evaluate our approach on the challenge dataset---derived from the MSP-Podcast corpus~\cite{LotfianBuildingNaturalistic1910b} and featuring naturalistic emotional speech---by addressing class imbalance.
To mitigate frequency bias while preserving diverse examples to learn emotions, we randomly subsample the majority classes in the training set to a maximum imbalance ratio of 8:1.
This choice balances the reduction of frequency bias present in the original training set with a maximum imbalance ratio of 26:1 with the need to preserve sufficient diversity for effective emotion learning.
Our final training subset comprises 49,248 utterances, 1,994 speakers, and 78 hours of recording.
For robust model selection, we create a balanced development set by sampling an equal number of instances per emotion category following the approach proposed by Härm \textit{et al.}~\cite{HarmTalTechSystems2406}, yielding 2608 utterances, 473 speakers, and 5 hours of audio.
We used the original test set for final evaluation.

Since test set transcripts are not provided, we generate consistent transcriptions across all splits using Whisper~\cite{RadfordRobustSpeech2212a}.
We used pretrained weights available at \url{https://huggingface.co/openai/whisper-large-v3}.

\section{Results and dicussion}

\subsection{Multimodal approach}

Initial experiments demonstrate the effectiveness of our multimodal approach compared to unimodal baselines.
As shown in Table~\ref{tab:multimodal_results}, our multimodal architecture achieves a macro F1 score of 0.3559 on the development set, outperforming both the audio-only and transcript-only models.
This suggests that audio and textual modalities provide complementary information for emotion recognition, with each modality capturing distinct emotional cues that contribute to better overall performance.


\begin{table}[b]
  \caption{Macro F1 Scores models on MSP-Podcast corpus.}
  \label{tab:multimodal_results}
  \centering
  \begin{tabular}{l|c|c}
    \toprule
    \textbf{Model} & \textbf{Dev Score} & \textbf{Test score}\\
    \midrule
    Official baseline (2025) & & 0.32 \\
    \midrule
    W2V-BERT 2.0 (audio only) & 0.2921 & \\
    DeBERTa V3 (transcript only) & 0.3174 & \\
    Multimodal baseline & 0.3559 & \\
    \bottomrule
  \end{tabular}
\end{table}

\subsection{Comparative pooling analysis}

As shown in Table~\ref{tab:results_pooling}, attention-based pooling strategies surpass static pooling (average/statistical) baselines when optimally parameterized, though improper configurations degrade performance due to their architectural complexity.
In what is following, we set parameter $p=256$.
Multi-head approaches are consistently configured with $n=1$, whereas alternative methods employ $n=2$.

MQMHA with $Q=2, H=2$ achieves the highest validation performance, indicating its capacity to disentangle multimodal emotion patterns effectively.
Due to time constraints, however, only the AS strategy was submitted for evaluation on the test set, despite MQMHA’s superior performance.
We attribute this decision to diminishing returns observed at higher query counts ($Q > 2$), where increased complexity risks overfitting without commensurate gains in generalization---an outcome aligned with sparse emotional label distributions.

\begin{table}[t]
  \caption{Macro F1 Scores on MSP-Podcast corpus using various pooling strategies.}
  \label{tab:results_pooling}
  \centering
  \begin{tabular}{l|c|c}
    \toprule
    \textbf{Pooling strategy} & \textbf{Dev score} & \textbf{Test score}\\
    \midrule
    Official baseline (2025) & & 0.3293 \\
    \midrule
    Average (\textit{Default}) & 0.3559 & \\
    Max  & 0.3418 & \\
    Statistics & 0.3623 & \\
    \midrule
    AS ($Q=1, H=1$) & 0.3884 & 0.3649 \\
    SA ($Q=2, H=1$) & 0.3416 & \\
    MHA ($Q=1, H=2$) & 0.3829 & \\
    \textbf{\boldmath MQMHA ($Q=2, H=2$)} & \textbf{\boldmath 0.3912} & \\
    MQMHA ($Q=2, H=4$) & 0.3675 & \\
    MQMHA ($Q=4, H=4$) & 0.3514 & \\
    \bottomrule
  \end{tabular}
\end{table}

\subsection{Localization of emotional cues}

\begin{figure}[h]
  \centering
  \includegraphics[width=\linewidth]{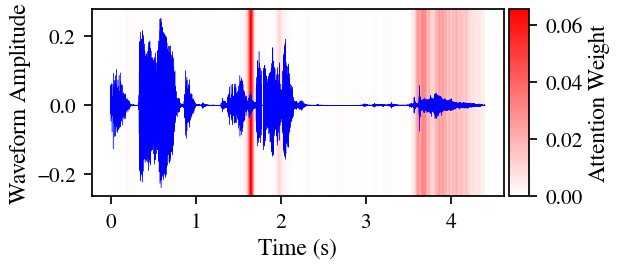}
  \caption{Attention heatmap on sample 5244\_0119.}
  \label{fig:attention_heatmap}
\end{figure}

While attention-driven classification gains suggest that models localize emotionally salient regions, it remains unclear whether such areas holistically capture emotional content.
We analyze attention heatmaps (from an AS pooling layer) and their temporal dynamics to scrutinize what the model learns---specifically, whether high-attention frames align with interpretable acoustic or linguistic cues, and how this informs emotion recognition efficacy.
Figure~\ref{fig:attention_heatmap} visualizes these weights for a sample utterance that was correctly classified as expressing sadness.
On the balanced development set, an averaged correlation of $\rho = 0.20 \pm 0.13$ between attention weights and audio energy suggests that the attention mechanism effectively focuses on emotionally salient regions rather than merely reflecting raw acoustic intensity.

\begin{figure}[h]
  \centering
  \includegraphics[width=\linewidth]{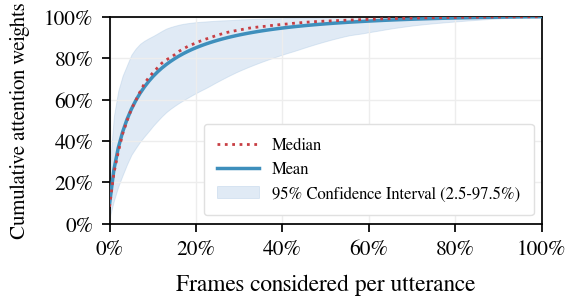}
  \caption{Attention weight distribution across frames.}
  \label{fig:attention_distribution}
\end{figure}

Figure~\ref{fig:attention_distribution} presents the aggregated cumulative distribution of attention weights across temporal frames, revealing the model's attention allocation patterns in sequential data.
The distribution exhibits a characteristic steep initial slope, where on average, 15\% of frames account for 80\% of the cumulative attention weight.
This highly concentrated distribution suggests that emotional cues are primarily localized within specific temporal regions rather than being uniformly distributed across the utterance.
The observed pattern follows a Pareto-like distribution, where a small subset of temporal frames captures the majority of the model's focus.
This finding suggests an efficient information extraction mechanism, where the attention layer successfully identifies and emphasizes the most emotionally salient segments.

\begin{figure}[h]
  \centering
  \includegraphics[width=1\linewidth]{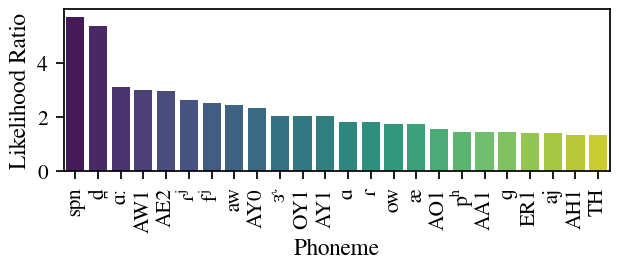}
  \caption{Attention phoneme salience as bayesian likelihood ratios.}
  \label{fig:phoneme_distribution}
\end{figure}

Figure~\ref{fig:phoneme_distribution} shows the results of our Bayesian likelihood analysis, which calculated the frequency of attended phonemes relative to their prior corpus frequency.
The analysis revealed systematic patterns in the most attention-grabbing phonemes at the utterance level, which were strongly correlated with phonetic prominence and acoustic salience.
We leveraged the phoneme annotations available in the MSP-Podcast corpus~\cite{LotfianBuildingNaturalistic1910b}.
The non-linguistic marker \texttt{spn} (spoken noise) was exceptionally overrepresented, indicating attention weights disproportionately prioritize non-speech vocalizations (e.g., breath sounds, laughter) during pooling.
Primary-stressed vowels (\texttt{AW1}, \texttt{AY1}) and diphthongs (\texttt{aw}) exhibited elevated ratios, consistent with their acoustic markedness (longer duration, higher intensity).
Secondary stress (\texttt{AE2}) also showed heightened salience, corroborating the role of syllabic prominence in perceptual weighting.
The attention mechanism’s prioritization of hyperarticulated vowels and non-canonical phones aligns with human perceptual strategies for decoding speech in noise~\cite{SchullerComputationalparalinguistics14, JuslinCommunicationemotions03}, reinforcing its biological plausibility in SER systems.

\section{Conclusions}

In this work, we introduced MQMHA to Speech Emotion Recognition, demonstrating its advantages over traditional attention mechanisms in the context of the Interspeech 2025 Speech Emotion Recognition Challenge in Naturalistic Conditions Challenge.
Our comprehensive evaluation showed that MQMHA outperforms Attentive Statistics and other pooling strategies while maintaining computational efficiency.
This advancement is integrated into a multimodal architecture combining W2V-BERT 2.0 and DeBERTa v3 into a lightweight multimodal architecture (less than 1B parameters), supported by considerations of class imbalance and gradual unfreezing strategies.
Through attention analysis, we revealed that attention-based pooling methods effectively identifie emotionally salient regions independently of signal energy, contributing to both performance improvement and interpretability in SER systems.

\ifinterspeechfinal
\section{Acknowledgements}
This publication was made possible by the use of the CEA-List FactoryIA supercomputer, financially supported by the Ile-de-France Regional Council; and has been performed in the scope of the OASEES Project, supported by the Commission of the European Communities \textit{/HORIZON, GA No.101092702}.
\else
\fi

\bibliographystyle{IEEEtran}
\bibliography{refs}

\end{document}